\documentclass[article,letterpaper,times,11ppt,listings-bw,microtype]{article}
\usepackage[utf8]{inputenc}
\usepackage[T1]{fontenc}
\usepackage{fixltx2e}
\usepackage{graphicx}
\usepackage{longtable}
\usepackage{float}
\usepackage{wrapfig}
\usepackage{soul}
\usepackage{textcomp}
\usepackage{marvosym}
\usepackage{wasysym}
\usepackage{latexsym}
\usepackage{amssymb}
\usepackage{hyperref}
\title{DUNE as an Example of Sustainable Open Source Scientific Software Development}
\author{Markus Blatt\thanks{Dr. Markus Blatt - HPC-Simulation-Software \& Service, Hans-Bunte-Str. 8-10, 69123 Heidelberg, Email: markus@dr-blatt.de}}
\date{\today}
\hypersetup{
  pdfkeywords={DUNE, sustainable software, scientific software, open source},
  pdfsubject={},
  pdfcreator={Emacs Org-mode version 7.8.11}}

\begin{document}

\maketitle

In this paper we describe how DUNE, an open source scientific software
framework, is developed. Having a sustainable software framework for
the solution of partial differential equations is the main driver of
DUNE's development. We take a look how DUNE strives to stay
sustainable software.
\section{The Cause for DUNE's Development}
\label{sec-1}

DUNE, the ``Distributed and Unified Numerics Environment''
\cite{DuneWeb,dune08-1,dune08-2,ISTL} is a modular
software toolbox for solving partial differential equations.
When initiating DUNE the rule in scientific software development was
that PhD students were given a legacy software and further extended it during
their PhD until it did fit their needs. The purpose of this custom
software was to produce publishable scientific results. The software
itself was only a (unwanted) by-product. As such it was neither
designed for maintainability nor well documented.

Even institutions that have an inhouse software or 
framework developed by a group of individual researchers suffer from
such development practices. This has produced many rather
unmaintainable software dinosaurs. They have a lot of hidden
undocumented features unknown to current developers. 
This leads to new researchers reimplementing functionality that they
need.

Often the developing institutions have a rather narrow application area
for their software. Therefore it is almost impossible to use the
software in a slightly different area as originally intended. This
again leads to researchers having to reimplement the wheel if they
want to switch the application area or even the underlying numerics.

This situation was a thorn in the eyes of the DUNE developers. At the
time of DUNE's inaugurating meeting 2002 they decided to attack
these problems. The developers come from different institutions
and have different (numerical) backgrounds. Quite a few of them had
their own legacy inhouse software when DUNE started, at that time mostly grid managers.
The purpose was to develop a unified interface to use them all
underneath for different purposes. This interface should be slim. That is
provide only absolutely necessary functionality. Nevertheless it
should be universally applicable to different numerical methods.
 In contrast to many older interfaces it should
be a fine grained interface in the sense that it is possible to access
individual entries in a container (e.g. vertices of a computational
grid or matrix entries) rather then just providing functions
that work on the whole containers. Given the rise of modern C++ with
templates this was possible without sacrificing computational efficiency.
\section{The Development Cycle: History, Current and Future}
\label{sec-2}

It took approximately three years for fixing the initial interface and
developing various prototypes. Some were full grids, others were
based on legacy code but only exposed the interface of a DUNE
grid. Another two years went by until we released the first stable
version and published the papers
\cite{dune08-1,dune08-2,ISTL} describing the interfaces. 

Already at that time the DUNE community was divided into three groups: the
so-called core developers that are allowed to vote on interface
decisions, normal DUNE developers that have write access to the source
code repositories, and other DUNE users that contribute with bug
report and patches. This division exists until now. Active patch
providers can become developers, if they find a core developer willing
to mentor and take responsibility for them. A DUNE developer can become
core developers with the support of the majority of the core
developers.

Currently DUNE users play a more and more important role collaborating
with the developers and we search for ways to raise their
participation. This will be done with more regular user meetings. The
first one was in 2010 and the next one will be in September 2013. This time
there will also be a developer meeting directly afterwards. This will
facilitate an even closer interaction between users and developers and
make sure that future developments fit the need of an even more varied
user base.

The main current aim is to establish a more forseeable
release cycle. We are planning to release roughly twice per year. This
allows users to focus their own development on the current stable
release and still be able to use most of the most recent features.
\section{Key Sustainability Drivers of DUNE}
\label{sec-3}
\subsection{Scientific Developers with Varied Backgrounds}
\label{sec-3-1}

The core developers of DUNE all come from various institutions,
have different scientific backgrounds (e.g. Navier Stokes, linear
elasticity, flow in porous media, multigrid methods), and use
different numerical methods. This makes sure that the software and its
interface is usable for various numerical methods.
\subsection{Open Source}
\label{sec-3-2}

From the beginning DUNE is available as open
source. Its license is GPL-2 with the so-called ``runtime
exception''. This exception allows the template code of the DUNE
modules to be used even in closed source projects. This was a
prerequisite of some of the initial DUNE developers because there are
German funding agencies that sponsor projects where both industry and
academia collaborate. Currently there are big companies that adopted
DUNE as their development platform due to its openness and even open
sourced some of their own products.
\subsection{Modularity}
\label{sec-3-3}

DUNE consists of various modules, each for its own purpose. Currently
there are seven core modules supported by the DUNE community and two
discretization modules developed by some DUNE developers. This lets
users choose just the functionality that they need or want.

When moving from one monolithic source tree to various modules, each
with its own source tree, the developers also implemented a rather
elaborate build system on top of autoconf and automake. This makes it
simple for users to create their own DUNE modules by simply calling a
script. The new module now has the same build system support as the core
DUNE modules and allows for code and build system reuse. 

Today there
are various other large frameworks based on DUNE: DUMUX, a
framework for multi-physics, multi-phase, and multi-domain simulation, \cite{DUMUXWEB},
and OPM, the open porous media initiative, \cite{OPMWEB},  that strives to build an
open source simulator suite for flow and transport in porous media for
the oil industry.
\subsection{Semi Open Development Model}
\label{sec-3-4}

The further development of DUNE is open. There are regular developer
and user meetings. Even the developer meetings are open to others
upon request. At the developer meetings important decisions, such as
interface changes, are made. The core developer, currently nine from
eight different institutions in Germany, the UK, and the USA, are
able to vote during decisions and a majority of them is needed for a
decision. Most decisions are discussed a priori on the mailing list or
the bug tracker. Smaller once are even decided there. This procedure
makes sure that only changes or additions are made that make sense to
and are really needed by the majority. Flaws in an initial proposal
are easily detected because of the diverse application areas of the
developers and users.

In recent years the cooperation of the DUNE users with the DUNE
development has increased. We receive various important bugfixes from
them. Due to the stability of the development branch of DUNE
experienced users often choose to use it in their everyday work and
development. This leads to many human testers and helps to further
improve the stability of DUNE.
\subsection{Code and Interface Reviews}
\label{sec-3-5}

Since the beginning DUNE was developed publically using modern version
control systems (VCS): first subversion and since 2013 git. All
committed patches are also send to a special mailing list. Most of the
core DUNE developers read this list and thus review the patches. This
often helps to detect bugs, possible problems with downstream user
modules, or not aproofed interface changes very
early in the development process. If developers want to change the
interface, then they have to provide a proposal that has to be
aproofed by the core developers at one of their meetings.
\subsection{Joint Development by Industry and Academia}
\label{sec-3-6}

Since 2011 one the core developers has become an entrepreneur and is now
providing DUNE contract work, and support. Commercial entities
participating in the DUNE development naturally have a long term interest in the
development. Often they have resources available for infrastructure
work (e.g. build system, testing, and usability features) and
naturally do quality management. This reduces the work
for scientific developers and lets them focus more on the science
side. In the end this will be advantageous for all.
\bibliographystyle{plain}
\bibliography{sustainable-dune}

\end{document}